\documentclass[aps,prd,twocolumn,showpacs,nofootinbib,amsmath,amssymb,amsfonts,showkeys]{revtex4}

\usepackage{graphicx}
\usepackage{bm}

\begin{document}

\title{Perturbed dark energy:\\classical scalar field versus tachyon}

\author{Olga Sergijenko}
 \email{olka@astro.franko.lviv.ua}
\author{Bohdan Novosyadlyj}
 \email{novos@astro.franko.lviv.ua}
\affiliation{Astronomical Observatory of 
Ivan Franko National University of Lviv, Kyryla i Methodia str., 8, Lviv, 79005, Ukraine}

\date{\today}

\begin{abstract}
The evolution of scalar linear perturbations is studied in gauge-invariant approach for 2-component models with nonrelativistic matter and minimally coupled scalar fields, the potentials of which were constructed for either constant dark energy equation of state (EoS) parameter $w$ or its adiabatic sound speed $c_a^2$ equal to zero. The numerical solutions show that such fields are almost smoothed out on subhorizon scales. However they cause the scale dependent suppression of the nonrelativistic mater density perturbations and the decay of gravitational potential, which can be used for choice of the dark energy model. We discuss 2 types of the Lagrangian: classical and tachyonic ones. As our results show, the fields with $w=const$ are almost indistinguishable, while for fields with $c_a^2=0$ the difference of dark energy effective sound speeds $c_s^2$, which is caused by the shape of Lagrangian, affects the evolution of perturbations significantly. We present also the transfer functions for both components.
\end{abstract}

\pacs{95.36.+x, 98.80.-k}
\keywords{cosmology: theory--dark energy--classical scalar field--tachyon field--evolution of scalar linear perturbations}

\maketitle

\section{Introduction}

Modern cosmological observations surely confirm that the expansion of the Universe is accelerated. Such phenomenon can be described using the models based on the Einstein equations with $\Lambda$-term \cite{lambda,wmap5}. Unfortunately, in this case several unsolved problems exist, e.g., the fine tuning and cosmic coincidence ones which are caused by the constant energy density corresponding to this fundamental quantity \cite{lambdaproblem}. Alternatively one can assume the existence of new mysterious essence -- dark energy. This repulsive component is regarded frequently as a scalar field, the Lagrangian of which can have either canonical kinetic term (classical field) \cite{clas,linder} or non-canonical one (e.g., tachyon \cite{tachyon}, k-essence \cite{kessence}).

It is often convenient to calculate the characteristics of the large-scale structure and CMB anisotropy in the phenomenological fluid framework, in which the dark energy model is defined by its density, equation of state (EoS) parameter and effective sound speed. The dynamical dark energy has been widely studied using both scalar field and fluid approaches on the background level as well as at the linear \cite{linear,c_a^2} and non-linear \cite{nonlinear} stages of perturbations growth for either minimal or non-minimal \cite{nonmin} coupling to other components. Moreover, the models unifying both dark matter and dark energy have been proposed \cite{udm}. For quintessential dark energy the EoS parameter $w$ is assumed to be larger than $-1$, the models with $w<-1$ are called phantom ones \cite{phant}. The phantom divide crossing is not allowed for single fields with the simplest Lagrangians, therefore in such case the specific single-field models as well as multi-field ones should be discussed (e.g., \cite{cros}).

The purpose of this paper is to perform the comparative analysis of 4 possible scalar field models of dark energy. This allows us to find the most sensitive to dark energy type features and thus to propose the simple tests for identification of the source accelerating the expansion of the Universe. As the theoretically motivated scalar field potentials \cite{modpot} frequently lead to the cosmological consequences contradicting the observations, we perform the reverse ingeneering, i.e., construct the potentials leading to the given dark energy properties (energy density, EoS parameter). We study the scalar fields with 2 shapes of Lagrangian: classical (Klein-Gordon) and tachyonic (Dirac-Born-Infeld) ones. For our research we have chosen the dark energy EoS parameter to be either constant or evolving in such way that the dark energy adiabatic sound speed $c_a^2$ equals 0. The reason for such choice is that in both cases the analytic solutions of the background field equations exist. The latter parametrization is especially interesting since for homogeneous fields it produces naturally the behavior mimicing the dust matter at early time and cosmological constant at very late epoch, so it can be regarded as possible candidate for so-called unified dark matter. All fields were studied assuming the same set of the best fitting cosmological parameters obtained by WMAP and other projects. Since the background dynamics in models with both Lagrangians is the same, the main attention should be paid to the linear stage of evolution of perturbations. On this level the degeneracy due to the Lagrangian type also exists, however the wide class of models becomes principally distinguishable. In this work we study the model of the Universe filled only with nonrelativistic matter and minimally coupled scalar field acting as dark energy. We focus on the influence of perturbed dark energy on matter perturbations and emphasize the effect of the inclusion of dust matter on evolution of dark energy inhomogeneities. The latter effect is commonly ignored due to the undetectability of dark energy perturbations. Nevertheless it should be also investigated as it might be important for better understanding of the gravitational interaction of both dark components and thus for discovering of the nature of dark energy. We discuss also the gravitational instability in the single-component models with $c_a^2=0$ in context of the unified dark matter.

The paper is organized as follows. In Section \ref{bckgr} we discuss briefly the background dynamics of our models. In Section \ref{evolsec} the evolution equations for the gauge-invariant perturbation variables are written and adiabatic initial conditions are set. In Section \ref{disc} we discuss the numerical solutions of the perturbations equations for the single fields with $c_a^2=0$, 2-component model of the Universe and build the transfer functions for both components. We propose also the characteristics which can be used for determination of the dark energy type. The conclusions can be found in Section \ref{concl}. In Appendix we present the evolution equations for the field variables.

\section{Cosmological background}\label{bckgr}

We consider the homogeneous and isotropic flat Universe with metric of 4-space 
$$ds^2=g_{ij} dx^i dx^j =a^2(\eta)(d\eta^2-\delta_{\alpha\beta} dx^{\alpha}dx^{\beta}),$$
where the factor $a(\eta)$ is the scale factor, normalized  to 1 at the current epoch $\eta_0$, $\eta$ is conformal time ($cdt=a(\eta)d\eta$). Here and below we put $c=1$, so the time variable $t\equiv x_0$  has the dimension of a length. The latin indices $i,\,j,\,...$ run from 0 to 3 and the greek ones over the spatial part of the metric: $\alpha,\, \beta,\,...$=1, 2, 3.

If the Universe is filled with non-relativistic matter (cold dark matter and baryons) and minimally coupled dark energy, the dynamics of its expansion is completely described by the Einstein equations 
\begin{eqnarray}
R_{ij}-{\frac{1}{2}}g_{ij}R=8\pi G \left(T_{ij}^{(m)}+T_{ij}^{(de)}\right),
\end{eqnarray}
where $R_{ij}$ is the Ricci tensor and $T_{ij}^{(m)}$, $T_{ij}^{(de)}$ -- energy-momentum tensors of matter $(m)$ and dark energy $(de)$. If these components interact only gravitationally then each of them satisfy the differential energy-momentum conservation law separately:
\begin{eqnarray}
T^{i\;\;(m,de)}_{j\;;i}=0
\end{eqnarray}
(here and below ``;'' denotes the covariant derivative with respect to the coordinate with given index in the space with metric $g_{ij}$). For the perfect fluid with density $\rho_{(m,de)}$ and pressure $p_{(m,de)}$, related by the equation of state $p_{(m,de)}=w_{(m,de)}\rho_{(m,de)}$, it gives
\begin{eqnarray}
\dot{\rho}_{(m,de)}=-3\frac{\dot a}{a} \rho_{(m,de)}(1+w_{(m,de)})
\end{eqnarray}
(here and below a dot denotes the derivative with respect to the conformal time: ``$\dot{\;\;}$''$\equiv d/d\eta$). The matter is considered to be non-relativistic, so  $w_m=0$ and $\rho_m=\rho_m^{(0)}a^{-3}$ (here and below ``0'' denotes the current values).

We assume the dark energy to be a scalar field with either Klein-Gordon (classical) Lagrangian 
\begin{eqnarray}
L_{clas}(X_{clas},\phi)=\frac{1}{2}\phi_{;i}\phi^{;i}-U(\phi)\label{Lclas}
\end{eqnarray}
or Dirac-Born-Infeld (tachyonic) one
\begin{eqnarray}
L_{tach}(X_{tach},\xi)=-\tilde{U}(\xi)\sqrt{1-\xi_{;i}\xi^{;i}},\label{Ltach}
\end{eqnarray}
where ${U}(\phi)$ and $\tilde{U}(\xi)$ are the field potentials defining the model, $X_{clas}=\phi_{;i}\phi^{;i}/2$ and $X_{tach}=\xi_{;i}\xi^{;i}/2$ are kinetic terms.

We postulate also that the background scalar fields are homogeneous ($\phi(\textbf{x},\eta)=\phi(\eta)$, $\xi(\textbf{x},\eta)=\xi(\eta)$), so their energy density and pressure depend only on time:
\begin{eqnarray}
&&\rho_{clas}=\frac{1}{2a^2}\dot{\phi}^{2}+{U}(\phi),\,\,\,\,\,
p_{clas}=\frac{1}{2a^2}\dot{\phi}^{2}-{U}(\phi),\\
&&\rho_{tach}=\frac{\tilde{U}(\xi)}{\sqrt{1-\dot{\xi}^2/a^2}},\,\,\,\,
p_{tach}=-\tilde{U}(\xi)\sqrt{1-\frac{\dot{\xi}^2}{a^2}}.
\end{eqnarray}
The evolution equations for field variables $\phi(\eta)$ and $\xi(\eta)$ are presented in Appendix \ref{ap}.

We specify the model of each field using 2 parameters: the EoS parameter $w_{de}\equiv p_{de}/\rho_{de}$ and the so-called adiabatic speed of sound
$c^2_a\equiv \dot{p}_{de}/\dot{\rho}_{de}$\footnote{The quantity $c_a^2$ doesn't play a role of the true sound speed of dark energy, it is rather a useful function, the form of which is similar to that of the adiabatic sound speed for barotropic EoS.}\cite{linder,unnikrishnan1,c_a^2,hu}, which satisfy the relation:
\begin{eqnarray}
 \dot{w}_{de}=3aH(1+w_{de})(w_{de}-c_a^2),
\end{eqnarray}
where $H=\dot{a}/{a^2}$ is the Hubble parameter (expansion rate) for any moment of conformal time $\eta$.
 Generally, the equation of state is defined by the Lagrangian as
\begin{eqnarray}
 w_{de}=\frac{L}{2X\frac{\partial L}{\partial X}-L},
\end{eqnarray}
however, different shapes of $L$ can lead to the same $w_{de}$ -- this is the well-known degeneracy on the background level.
 The scalar field evolution equations have the analytical solutions for two cases:
\begin{itemize}
 \item $w=const$: $c_a^2=w,\,\,\,\,\rho_{de}(a)=\rho_{de}^{(0)}a^{-3(1+w)}$ and
\item $c^2_a=0$: $w(a)=w_0a^3/(1+w_0-w_0a^3),\,\,\,\,\rho_{de}(a)=\rho_{de}^{(0)}\left[(1+w_0)a^{-3}-w_0\right]$.
\end{itemize}
Here and below we omit index $de$ denoting both -- classical and tachyonic -- scalar fields for $w_{de}$.

Note that the fields with $c_a^2=0$ at $a\ll 1$ behave as the dust matter and at $a\rightarrow\infty$ they mimic the cosmological constant. This fact suggests that they can also be studied as the possible candidates for the so-called unified component, or unified dark matter, describing the early matter dominated stage and current $\Lambda$ or dark energy dominated one by introducing a single new fluid, e.g., Chaplygin gas \cite{udm}.

In this case the expansion rate and fields evolve as follows:
\begin{eqnarray}
 H=H_0a^{-\frac{3}{2}}\sqrt{1+w_0-w_0a^3},&&\\
\phi(a)-\phi_0=\pm\frac{1}{2\sqrt{6\pi G}}&&\nonumber\\
\times\ln\left(\frac{\sqrt{1+w_0(1-a^3)}-\sqrt{1+w_0}}{\sqrt{1+w_0(1-a^3)}+\sqrt{1+w_0}}
\frac{1+\sqrt{1+w_0}}{1-\sqrt{1+w_0}}\right),&&\\
\xi(a)-\xi_0=\pm\frac{2}{3H_0\sqrt{-w_0}}\left[\arctan\left(\sqrt{-\frac{w_0a^3}{1+w_0}}\right)\right.&&\nonumber\\
\left.-\arctan\left(\sqrt{-\frac{w_0}{1+w_0}}\right)\right].&&
\end{eqnarray}

For the 2-component model the Hubble parameter $H(a)$ behaves as:
\begin{eqnarray}
H=H_0a^{-\frac{3}{2}}\sqrt{1-\Omega_{de}+\Omega_{de}a^{-3w}}
\end{eqnarray}
for $w=const$ and 
\begin{eqnarray}
H=H_0a^{-\frac{3}{2}}\sqrt{1+\Omega_{de}w_0-\Omega_{de}w_0a^3}
\end{eqnarray}
for $c^2_a=0$.

The temporal dependences of scalar fields and their potentials are following:
\begin{eqnarray}
&&\phi(a)-\phi_0=\pm\int_1^a\frac{da'\sqrt{\rho_{de}(a')(1+w(a'))}}{a'H(a')},\\
&&U(a)=\frac{\rho_{de}(a)\left[1-w(a)\right]}{2}
\end{eqnarray}
for the classical Lagrangian and
\begin{eqnarray}
&&\xi(a)-\xi_0=\pm\int_1^a\frac{da'\sqrt{1+w(a')}}{a'H(a')},\\
&&\tilde{U}(a)=\rho_{de}(a)\sqrt{-w(a)}
\end{eqnarray}
for the tachyonic one.

In this paper we study the classical and tachyonic scalar fields with potentials constructed for $w=const$ and $c_a^2=0$ using the same set of the best fitting cosmological parameters from \cite{wmap5} ($\Omega_{de}=0.722$, $w=w_0=-0.972$, $\Omega_m=0.278$, $h=0.697$). For all models there are 2 independent solutions for the field: the growing one (sign +) and the decaying one (sign -). Therefore 2 symmetrical with respect to either $\phi-\phi_0$ or $\xi-\xi_0$ potentials exist in each case \cite{olka2008,novos2008}. However, the physical consequences of both these solutions are the same, so from now we restrict ourselves only to the growing one.

The analysis of dynamics of the Universe expansion for fields with $w=const$ and $c_a^2=0$ was presented in \cite{olka2008,novos2008}. It doesn't depend on the scalar field Lagrangian and doesn't allow us to distinguish surely such models of scalar fields. So in order to propose the test for choice of the dark energy type which would be more adequate to observations we should study at least the linear stage of the evolution of scalar perturbations.

\section{Evolution of scalar linear perturbations}\label{evolsec}
For derivation of the evolution equations for scalar linear perturbations it is convenient to use the conformal-Newtonian gauge with space-time metric \cite{ma_bertschinger,novos2007}
\begin{eqnarray}
ds^2&=&a^2(\eta)[(1+2\Psi(\textbf{x},\eta))d\eta^2\nonumber\\
&&-(1+2\Phi(\textbf{x},\eta))\delta_{\alpha\beta}dx^{\alpha}dx^{\beta}],
\end{eqnarray}
where $\Psi(\textbf{x},\eta)$ and $\Phi(\textbf{x},\eta)$ are gauge-invariant metric perturbations called Bardeen's potentials \cite{bardeen}, which in the case of zero proper anisotropy of medium (as for the dust matter and scalar fields) have equal absolute values and opposite signs: $\Psi(\textbf{x},\eta)=-\Phi(\textbf{x},\eta)$ \cite{kodama_sasaki}.
In the linear perturbation theory it is convenient to perform the Fourrier transform of all spatially-dependent variables, so the equations will be written for the corresponding Fourier amplitudes of the metric ($\Psi(k,\eta)$), matter density and velocity perturbations ($\delta^{(m)}(k,\eta)$,
$V^{(m)}(k,\eta)$) as well as the scalar field  perturbations ($\delta^{(de)}(k,\eta)$, $V^{(de)}(k,\eta)$, $\delta{\phi}(k,\eta)$, $\delta{\xi}(k,\eta)$) (here $k$ is wave number). These variables are gauge-invariant \cite{kodama_sasaki,durrer,novos2007}. The energy density and velocity perturbations of dark energy, $\delta^{(de)}$ and $V^{(de)}$, are connected with the perturbation of field variables $\delta{\phi}$, $\delta{\xi}$ in the following way:
\begin{eqnarray}
&&\delta^{(clas)}=(1+w)\left(\frac{\dot{\delta{\phi}}}{\dot{\phi}}-\Psi+\frac{a^2\delta{\phi}}{\dot{\phi}^2}\frac{d{U}}{d\phi}\right),\\
&&V^{(clas)}=\frac{k\delta{\phi}}{\dot{\phi}},\\
&&\delta^{(tach)}=-\frac{1+w}{w}\left(\frac{\dot{\delta\xi}}{\dot{\xi}}-\Psi\right)+\frac{d\tilde{U}}{\tilde{U}d\xi}\delta{\xi},\\
&&V^{(tach)}=\frac{k\delta{\xi}}{\dot{\xi}}.
\end{eqnarray}

Other non-vanishing gauge-invariant perturbations of scalar field are isotropic pressure perturbation 
\begin{eqnarray}
&&\pi_L^{(clas)}=\frac{1+w}{w}\left(\frac{\dot{\delta{\phi}}}{\dot{\phi}}-\Psi-\frac{a^2\delta{\phi}}{\dot{\phi}^2}\frac{d{U}}{d\phi}\right),\\
&&\pi_L^{(tach)}=\frac{1+w}{w}\left(\frac{\dot{\delta\xi}}{\dot{\xi}}-\Psi\right)+\frac{d\tilde{U}}{\tilde{U}d\xi}\delta{\xi}
\end{eqnarray}
and intrinsic entropy
\begin{eqnarray}
\Gamma^{(de)}=\pi_L^{(de)}-\frac{c^2_a}{w}\delta^{(de)}.
\end{eqnarray}
The density perturbation of any component in the conformal-Newtonian gauge $D_s\equiv\delta$, which is gauge-invariant variable, is related to the other
gauge-invariant variables of density perturbations $D$ and $D_g$ as:
\begin{eqnarray}
&&D=D_g+3(1+w)\left(\Psi+\frac{\dot{a}}{a}\frac{V}{k}\right)\nonumber\\
&&=D_s+3(1+w)\frac{\dot{a}}{a}\frac{V}{k},\label{ddgds}
\end{eqnarray}
where $D_s$, $D$, $D_g$ and $V$ correspond to either $m$- or $de$-component. Here $D_g$ is the density perturbation in the frame of vanishing curvature fluctuations and $D$ corresponds to the rest frame, i.e., the frame in which the 4-velocity is orthogonal to constant time hypersurface.

For the scalar fields the intrinsic entropy perturbation is defined as \cite{hu,c_s^2,c_s^2_sp}:
\begin{eqnarray}
&&w\Gamma^{(de)}=\left(c_s^2-c_a^2\right)D^{(de)},
\end{eqnarray}
where $c_s^2$ is the dark energy effective (rest frame) sound speed:
\begin{eqnarray}
 c_s^2=\frac{\delta p^{(rf)}}{\delta\rho^{(rf)}}=\frac{1}{2X\frac{\partial^2L}{\partial X^2}+\frac{\partial L}{\partial X}}\frac{\partial L}{\partial X}.
\end{eqnarray}
For Lagrangians with canonical kinetic term it is always 1, for tachyonic ones $c_s^2=-w$.

For further study it is convenient to use the evolution equations for gauge-invariant density and velocity perturbation variables. The corresponding equations for perturbations of field variables can be found in Appendix \ref{ap}.

It should be noted that the principal degeneracy due to the Lagrangian shape exists also in the linear theory, i.e., different Lagrangians can lead to the same value of $c_s^2$ \cite{unnikrishnan1}. However, the analysis of this stage of evolution of perturbations is still important since it removes partially the degeneracy existing on the background level.

\subsection{Evolution equations}

The linearised Einstein equations for gauge-invariant perturbations of metric, density and velocity are
\begin{eqnarray}
&&\dot{\Psi}+aH\Psi-\frac{4\pi Ga^2}{k}\left[\rho_mV^{(m)}+\rho_{de}(1+w)\nonumber\right.\\
&&\left.\times V^{(de)}\right]=0,\label{evol1}\\
&&\dot{V}^{(m)}+aHV^{(m)}-k\Psi=0,\label{evol2}\\
&&\dot{D_g}^{(m)}+kV^{(m)}=0,\label{evol3}
\end{eqnarray}
\begin{eqnarray}
&&\dot{V}^{(de)}+aH(1-3c_s^2)V^{(de)}-k(1+3c_s^2)\Psi-\frac{c_s^2k}{1+w}\nonumber\\
&&\times D_g^{(de)}=0,\label{evol4}\\
&&\dot{D_g}^{(de)}+3(c_s^2-w)aHD_g^{(de)}+(1+w)\left[k+\frac{9}{k}a^2H^2\left(c_s^2\right.\right.\nonumber\\
&&\left.\left.-c_a^2\right)\right]V^{(de)}+9aH(1+w)\left(c_s^2-c_a^2\right)\Psi=0.\label{evol5}
\end{eqnarray}
Here and below $\rho$ corresponds to the background density of each component.

So, as we see, for each parametrization of EoS the shape of Lagrangian affects the evolution of perturbations only through the effective sound speed of dark energy.
In $w=const$-case $c_{s(clas)}^{2}=1$, $c_{s(tach)}^{2}=-w$, that in principle allows us to distinguish both Lagrangians, however the difference isn't large (for $w$ close to $-1$ -- as it has been estimated on basis of the observable data \cite{wmap5}) and suggests the similarity of solutions for both fields. In $c_a^2=0$-case $c_{s(clas)}^{2}=1=const$ but
$c_{s(tach)}^{2}=-w(a)\neq const$, so the behavior of perturbations in classical field and tachyon with the same cosmological parameters should be really different.

Therefore, in each case we have the system of 5 first-order ordinary differential equations for 5 unknown functions $\Psi(k,a)$, $D_g^{(m)}(k,a)$, $V^{(m)}(k,a)$, $D_g^{(de)}(k,a)$ and $V^{(de)}(k,a)$. The following constraint equation is also satisfied:
\begin{eqnarray}
 -k^2\Psi=4\pi Ga^2\left(\rho_mD^{(m)}+\rho_{de}D^{(de)}\right).\label{constrainteq}
\end{eqnarray}

\subsection{Initial conditions}
It is known that the observable large-scale structure has grown from the small adiabatic perturbations generated in the early Universe. Since the density of the $w=const$-fields is negligible at the early epoch ($a\ll 1$) and $c_a^2=0$-fields mimic dust matter, all our models are initially matter-dominated. In such case the growing mode of adiabatic perturbations corresponds to $\Psi_{init}=const$. So, here we specify the adiabatic initial conditions for the growing mode of perturbations \cite{doran,kodama_sasaki,durrer}:
\begin{eqnarray}
&&{V^{(de)}}_{init}=\frac{2}{3}\frac{k}{H_0}\frac{\Psi_{init}}{\sqrt{1-\Omega_{de}}}\sqrt{a_{init}},\\
&&{D_g^{(de)}}_{init}=-5(1+w)\Psi_{init},\\
&&{V^{(m)}}_{init}=\frac{2}{3}\frac{k}{H_0}\frac{\Psi_{init}}{\sqrt{1-\Omega_{de}}}\sqrt{a_{init}},\\
&&{D_g^{(m)}}_{init}=-5\Psi_{init}
\end{eqnarray}
for $w=const$ and 
\begin{eqnarray}
&&{V^{(de)}}_{init}=\frac{2}{3}\frac{k}{H_0}\frac{\Psi_{init}}{\sqrt{1+\Omega_{de}w_0}}\sqrt{a_{init}},\\
&&{D_g^{(de)}}_{init}=-5\Psi_{init},\\
&&{V^{(m)}}_{init}=\frac{2}{3}\frac{k}{H_0}\frac{\Psi_{init}}{\sqrt{1+\Omega_{de}w_0}}\sqrt{a_{init}},
\end{eqnarray}
\begin{eqnarray}
&&{D_g^{(m)}}_{init}=-5\Psi_{init}
\end{eqnarray}
for $c^2_a=0$.

\begin{figure} 
\centerline{\includegraphics[width=6.5cm]{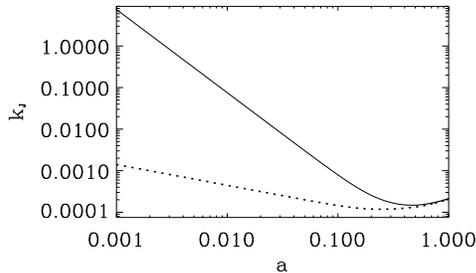}}
\caption{The Jeans scale for the Universe filled only with either classical (dotted line) or tachyonic (solid) field with $c_a^2=0$.}
\label{fig0}
\end{figure}

In all cases we have integrated numerically the systems of 1-order equations for these initial conditions using the publicly available code DVERK\footnote{It was created by T.E. Hull, W.H.Enright, K.R. Jackson in 1976 and is available at http://www.cs.toronto.edu/NA/dverk.f.gz}. We assumed $\Psi_{init}=-1$, $a_{init}=10^{-10}$. Such early initial time is chosen only in order to separate properly the growing adiabatic mode while our purpose is to analyze its behavior after recombination epoch, when the Universe can be effectively described by the 2-component (dust matter plus dark energy) model. The evolution of perturbations is scale dependent, so we performed calculations for the range of $k$ from $0.0001$ to $0.1$ Mpc$^{-1}$. The reason for such choice is that at present epoch larger scales are super-horizon while smaller ones are affected by the non-linear effects.

\section{Results and discussion}\label{disc}

\subsection{Single-component system: perturbed scalar field}
First of all let us discuss the Universe filled only with a scalar fields. Such models with $c_a^2=0$, as the background dynamics suggests, could probably be considered as simple candidates for the unified dark matter. Therefore we are going to analyze their clustering properties.
For such purpose it is useful to obtain from the first-order equations (\ref{evol4})-(\ref{evol5}) using (\ref{constrainteq}) a second-order one for the density perturbation:
\begin{eqnarray}
 &&D''+\left(\frac{3}{2a}-\frac{15}{2a}\frac{w_0a^3}{1+w_0-w_0a^3}\right)D'\nonumber\\
&&+\left[\frac{c_s^2k^2}{H_0^2a(1+w_0-w_0a^3)}+\frac{9}{2a^2}\left(\frac{w_0a^3}{1+w_0-w_0a^3}\right)^2\right.\nonumber
\end{eqnarray}
\begin{eqnarray}
&&\left.-\frac{12}{a^2}\frac{w_0a^3}{1+w_0-w_0a^3}-\frac{3}{2a^2}\right]D=0.\label{01}
\end{eqnarray}\begin{figure*}
\centerline{\includegraphics[width=14cm]{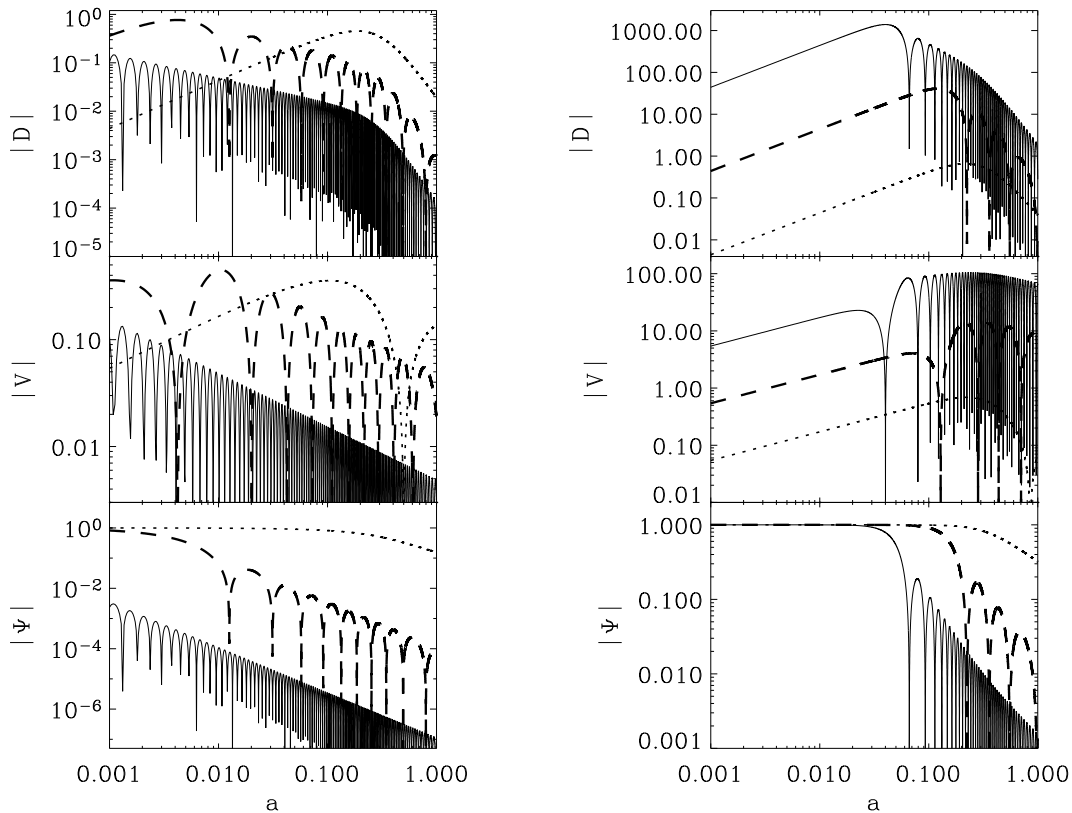}}
\caption{The evolution of the density (top) and velocity perturbations (medium) in a single-component model with $c_a^2=0$ is presented for scales $k=0.0001$ (dotted line), $0.001$ (dashed), $0.01$ Mpc$^{-1}$ (solid). Bottom: the corresponding gravitational potentials. Classical field -- left column, tachyon -- right one.}\label{fig1}
\end{figure*}
The above equation could be written in the form:
\begin{equation}
u''+\omega^2u=0,\label{uom}
\end{equation}
where $u$ stands for $Da^{\frac{3}{4}}\left(1+w_0-w_0a^3\right)^{\frac{5}{4}}$.
The coefficient $\omega^2$ is generally a function of time and scale:
\begin{eqnarray*}
&&\omega^2=\frac{c_s^2k^2}{H_0^2a(1+w_0-w_0a^3)}-\frac{21}{16a^2}+\frac{9}{8a^2}\\
&&\times\frac{w_0a^3}{1+w_0-w_0a^3}+\frac{27}{16a^2}\left(\frac{w_0a^3}{1+w_0-w_0a^3}\right)^2.
\end{eqnarray*}
 However, in the neighborhood of any given point $a$ it can be regarded as constant with respect to $a$, thus it depends only on the wave number $k$. The condition $\omega^2(k)=0$ specifies for each $a$ the Jeans scale, under which the perturbations at this $a$ are stable (oscillatory) while above they are gravitationally unstable, i.e., can grow or decay:
\begin{eqnarray}
&&k_J^2=\frac{H_0^2}{c_s^2}\left[\frac{21}{16a}(1+w_0-w_0a^3)-\frac{9w_0a^2}{8}\right.\nonumber\\
&&\left.-\frac{27w_0a^2}{16}\frac{w_0a^3}{1+w_0-w_0a^3}\right].
\end{eqnarray}
The Jeans scale for the values of $a$ from 0.001 to 1 is shown in Fig. \ref{fig0}. For each field the values of $k$ below the corresponding curve define the perturbations which can cluster for the scale factors close to $a$. As we see, the perturbations with $k\lesssim0.0001$ Mpc$^{-1}$ are unstable all the time up to the current epoch, however they are still super-horizon, so irrelevant for the choice of model best fitting to the data. The perturbations with subhorizon scales could be unstable at early stages, but should oscillate at late times.

In Fig. \ref{fig1} the evolution of density, velocity and metric perturbations is shown. We see that generally the behavior of subhorizon modes is oscillatory. For classical Lagrangian the gravitational potential decays very quickly, so that at $a=0.001$ for $k=0.01$ Mpc$^{-1}$ (horizon entry at $a\approx0.0004$) it is already almost 0 while for $k=0.001$ Mpc$^{-1}$ (horizon entry at $a\approx0.04$) it is approximately $-0.8$ and starts to oscillate just before the horizon entry. For tachyonic Lagrangian the potential $\Psi$ at early stages is constant, then it begins to decay. Note that the perturbations of tachyon with larger $k$ change the character of the temporal dependence earlier (for $0.001$ Mpc$^{-1}$ at $a\approx 0.09$, for $0.01$ Mpc$^{-1}$ at $a\approx 0.03$), but the amplitude of the first positive peak is almost the same ($\approx 0.2$). The gravitational potential for the super-horizon mode with $k=0.0001$ Mpc$^{-1}$ decays monotonously, as expected.

The rest frame pressure perturbations $\delta p^{(rf)}=c_s^2\delta\rho^{(rf)}$ for $c_s^2>0$ have the same sign as $\delta\rho^{(rf)}$, therefore the nature of the scalar field oscillations is similar to that of the acoustic ones in baryon-photon plasma.

The presented in Fig. \ref{fig1} evolution of density perturbations (related to $\Psi$ as $D=-k^2\Psi/(4\pi Ga^2\rho)$) confirms such scenario. We see that for the same scale the magnitude of perturbation for the tachyon field is higher than for the classical one and that at late times all subhorizon modes are smoothed out in oscillatory manner. The amplitudes of velocity perturbations oscillations at first grow slightly but then start to decay.

The difference between the studied models is caused by the behavior of the effective sound speed. The perturbations of the classical field always propagate with the speed of light ($c_s^2=1$), so their evolution isn't close to that of dust matter at any stage of the growth. In contrary, the perturbations of tachyon propagate with the speed variable in time: $c_s^2=-w_0a^3/(1+w_0-w_0a^3)$. At early stages it is negligible, so such field mimics the dust matter not only at the background but at the linear stage of evolution of the perturbation too. Later, the effective sound speed grows and tends to the speed of light at infinity (as well as the equation of state parameter tends to $-1$). The behavior of the field (and metric) perturbations begins then to differ from the dust matter one and soon becomes oscillatory with decreasing amplitude.

Such features of the studied fields mean that they couldn't play the role of unified dark matter.

The model with a single $w=const$-field corresponds to the quasi-de Sitter Universe. In this case the perturbation equation takes the form:
\begin{eqnarray}
&&D''+\left(\frac{3}{2a}-\frac{9w}{2a}\right)D'+\left[\frac{c_s^2k^2}{H_0^2}a^{3w-1}\right.\nonumber\\
&&\left.+\frac{1}{a^2}\left(\frac{9w^2}{2}-3w-\frac{3}{2}\right)\right]D=0\label{w2}
\end{eqnarray}
and has the analytical solution:
\begin{eqnarray}
&&D=a^{\frac{9w-1}{4}}Z_{-\frac{1}{2}\frac{5+3w}{1+3w}}\left(\frac{2}{1+3w}\frac{c_sk}{H_0}a^{\frac{1+3w}{2}}\right),\label{w2s}
\end{eqnarray}
where $Z_{\nu}(x)=C_1J_{\nu}(x)+C_2Y_{\nu}(x)$ and $J_{\nu}(x), Y_{\nu}(x)$ are Bessel functions of first and second kind, $C_1, C_2$ -- arbitrary constants. For $a\ll1$ the density perturbations behave as $\tilde{C}_1a^{-\frac{3}{2}(1-w)}+\tilde{C}_2a^{\frac{1+3w}{4}}$ -- both modes decay for $w<-1/3$. In far future ($a\rightarrow\infty$) the amplitude of $D$ will decay as $a^{\frac{3w-1}{2}}$. Interestingly, the current oscillatory behavior of perturbations is clear also without analytical solution (\ref{w2s}). Really, if we put equation (\ref{w2}) in form of (\ref{uom}), the quantity
\begin{eqnarray*}
\omega^2=\frac{c_s^2k^2}{H_0^2}a^{3w-1}-\frac{9w^2}{16a^2}-\frac{15w}{8a^2}-\frac{21}{16a^2}
\end{eqnarray*}
is larger than $0$ up to present moment for all scales of interest.

\subsection{2-component system: dust matter and dark energy}
\begin{figure}
\centerline{\includegraphics[width=5cm]{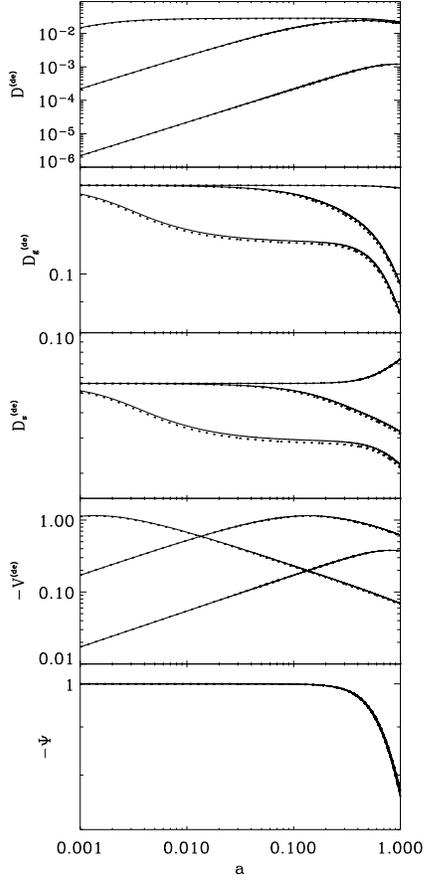}}
\caption{The evolution of variables $D^{(de)}$ ($k=0.01,\, 0.001$ and $0.0001$ Mpc$^{-1}$ from top to bottom), $D_g^{(de)}$, $D_s^{(de)}$, $\Psi$ (corresponding scales from bottom to top) and $V^{(de)}$ ($k=0.001,\, 0.0001$ and $0.01$ Mpc$^{-1}$ from top to bottom at $a\approx1$). Classical field with $w=const$ -- dotted line, tachyonic one -- solid. The curves for both fields are very close and practically indistinguishable.}
\label{fig2}
\end{figure}
Now we return to the minimally coupled 2-component model. The general conclusion is that in the studied case matter clusters while dark energy tends to homogeneity.

For the dark energy with $w=const$ the simple conclusion, that the behavior of scalar linear perturbations in model with the tachyonic field should be similar to that in the model with the corresponding classical field, is valid, as the numerical analysis has shown \cite{kul2008}. In plots presented in Fig. \ref{fig2} it is hard to distinguish the curves for the classical field and tachyon, so in this case we practically can't choose the Lagrangian preferred by observations. For scale $k=0.0001$ Mpc$^{-1}$ the behavior of dark energy density perturbations is different: $D_g^{(de)}$ remains almost constant while $D^{(de)}$ and $D_s^{(de)}$ grow, but this depends only on gauge choice for super-horizon modes.

\begin{figure*}
\centerline{\includegraphics[width=14cm]{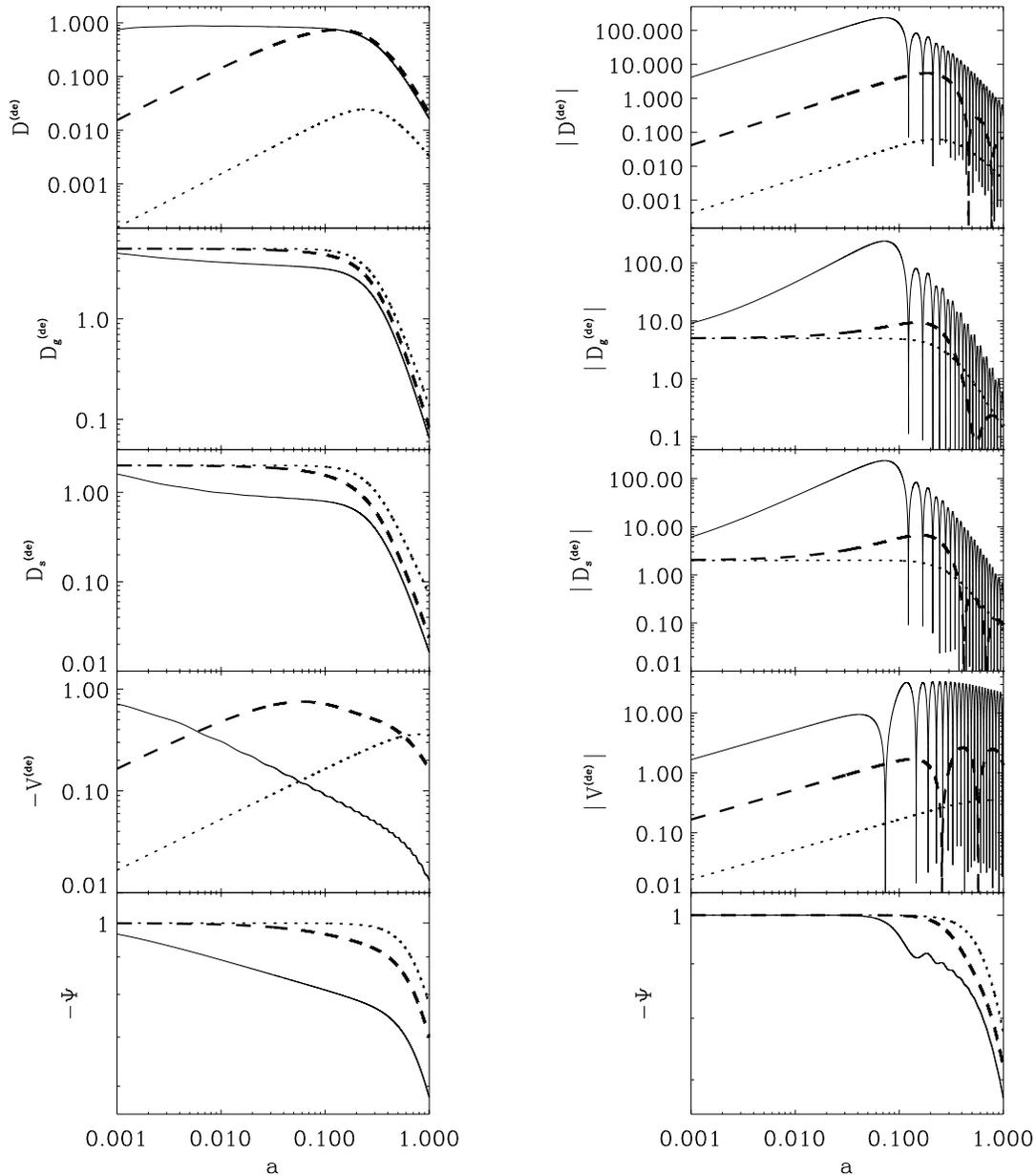}}
\caption{The evolution of density perturbations $D^{(de)}$, $D_g^{(de)}$, $D_s^{(de)}$, velocity ones $V^{(de)}$ and gravitational potential with scales $k=0.0001$ (dotted line), $0.001$ (dashed) and $0.01$ Mpc$^{-1}$ (solid) for classical (left) and tachyonic (right) fields with $c_a^2=0$.}\label{fig3}
\end{figure*}

In the $c_a^2=0$-case both fields are almost smoothed out on subhorizon scales at present and future epochs while on superhorizon ones they do not grow significantly. However, the difference of the effective sound speed behavior has sufficient imprint in the evolution of inhomogeneities, so, as it can be seen in Fig. \ref{fig3}, the perturbations of the classical field at early stages grow slowly but decay after the horizon entry while the perturbations of tachyon at first grow significantly and then begin to oscillate. The oscillation amplitudes decrease all the time for the density perturbations or increase at early time while decrease at the current epoch for the velocity ones. It should be noted that small oscillations along the averaged solutions for $V^{(de)}$ and $D^{(de)}$ are also present in classical field, however their amplitudes are highly subdominant comparing to the corresponding mean quantities. Interestingly, for perturbed tachyonic field at scale $k=0.001$ Mpc$^{-1}$ $D_g$ remains positive. This suggests that the small-scale perturbations of tachyon oscillate along the mean curves too, but in this case the averaged values are much smaller (almost negligible) comparing to the amplitudes of oscillations.

The influence of dust matter component on dark energy perturbations is described by the equation:
\begin{equation}
 \ddot{D}^{(de)}+\mathcal{A}\dot{D}^{(de)}+\mathcal{B}D^{(de)}+\mathcal{S}=0,
\end{equation}
where
\begin{widetext}
\begin{eqnarray}
\mathcal{A}&=&-\left(3w+\frac{k^2(3w-3c_a^2-1)+12\pi Ga^2\rho_m(3w-3c_a^2-2)}{k^2+12\pi Ga^2\rho_m}\right)aH,\\
\mathcal{B}&=&c_s^2k^2+3wa^2H^2\frac{k^2(3w-3c_a^2-1)+12\pi Ga^2\rho_m(3w-3c_a^2-2)}{k^2+12\pi Ga^2\rho_m}-3w(aH)\dot{}-9a^2H^2(1+w)(w-c_a^2)\nonumber\\
&&-4\pi Ga^2\rho_{de}(1+w)+12\pi Ga^2\rho_mc_s^2,\\
\mathcal{S}&=&-\left[D^{(m)}+3aH\left(3w-3c_a^2-2-\frac{k^2(3w-3c_a^2-1)+12\pi Ga^2\rho_m(3w-3c_a^2-2)}{k^2+12\pi Ga^2\rho_m}\right)\frac{V^{(m)}}{k}\right]\nonumber\\
&&\times4\pi Ga^2\rho_m(1+w).
\end{eqnarray}
\end{widetext}
For $\rho_m=0$ it takes the form of (\ref{01}). It can be easily seen that for the same values of cosmological parameters the non-zero density of nonrelativistic matter changes the coefficient $\mathcal{B}$ and thus the character of evolution of the dark energy perturbations more drastically for fields with larger effective sound speed.

In the bottom panel of Fig. \ref{fig3} the evolution of $-\Psi$ is shown for models with $c_a^2=0$ for the scales of perturbations $k=0.0001$, $0.001$ and $0.01$ Mpc$^{-1}$. It can be easily seen that in this case the strong model and scale dependences of $\Psi$ exist.

\begin{figure}
\centerline{\includegraphics[width=5cm]{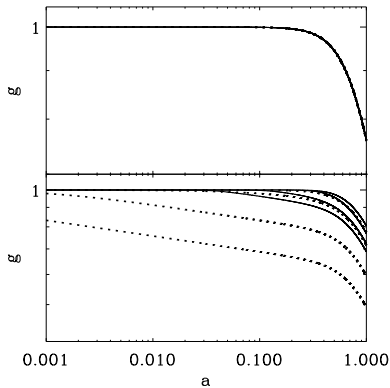}}
\caption{The evolution of $g$ for linear perturbations with scales $k=0.0001$, $0.001$, $0.01$ and $0.1$ Mpc$^{-1}$ (from top to bottom in each panel) in the models with non-relativistic matter and dark energy with $w=const$ (top) or $c_a^2=0$ (bottom). Classical field -- dotted line, tachyon -- solid line. Both axes are logarithmic.}\label{fig4}
\end{figure}

Generally, at present epoch the growth of matter density perturbations (described by the growth function $g$) is suppressed and -- unlike $\Lambda$CDM-case -- such suppression is scale dependent. The growth function is defined as follows:
\begin{equation}
 g\equiv D^{(m)}a_{init}/D^{(m)}_{init}a.
\end{equation}
It is convenient to compute this quantity using the solutions of (\ref{evol1})-(\ref{evol5}) and taking into account (\ref{ddgds}).
In the top panel of Fig. \ref{fig4} the temporal evolution of $g$ is shown for the fields with $w=const$. We see that all curves practically overlap. It means that the scale dependences are very weak and confirms the absence of substantial difference between classical and tachyonic dark energy.
The temporal dependences of $g$ for the fields with $c_a^2=0$ are shown for different scales in the bottom panel of Fig. \ref{fig4}. The suppression caused by the classical field is stronger than the corresponding suppression caused by tachyon. The behavior of the growth function can be used for choice of the adequate model of dark energy because it is connected with the dust matter transfer function and thus with the observable matter density fluctuations power spectrum.

The faster decay of $\Psi$ and $g$, caused by the classical field, is clear. From the equation
\begin{eqnarray}
&&\ddot{\Psi}+3\frac{\dot{a}}{a}\dot{\Psi}+\left[2\frac{\ddot{a}}{a}-\left(\frac{\dot{a}}{a}\right)^2\right]\Psi=4\pi Ga^2\rho_{de}\left(c_s^2D^{(de)}\right.\nonumber\\
&&\left.-3\frac{\dot{a}}{a}c_a^2(1+w)\frac{V^{(de)}}{k}\right)
\end{eqnarray}
we see that for models with the same EoS parameter and adiabatic sound speed of dark energy the influence of its perturbations on the gravitational potential and thus on dust matter inhomogeneities growth increases with the values of the dark energy effective sound speed. As for classical field $c_s^2$ always equals $1$ and for tachyon it initially is close to $0$ while at the current epoch it equals $-w_0$, which is smaller than $1$, the perturbations of the former field affect the matter and metric perturbations more. The scale dependences of $\Psi$ and $g$, caused by the dark energy perturbations and therefore absent in $\Lambda$CDM-cosmology, are also expected to be stronger for classical Lagrangian.

It should be noted that the oscillatory behavior of perturbations of the tachyonic field leads to small oscillations along the averaged solution for $\Psi$, but their amplitudes are subdominant comparing to the decaying mean dependence.

\begin{figure*}
\centerline{\includegraphics[width=14cm]{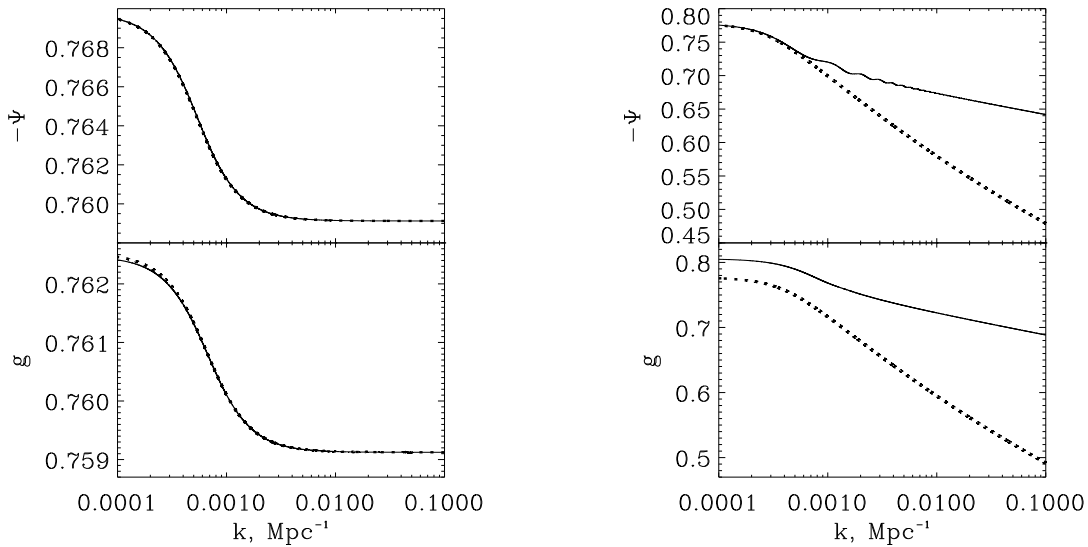}}
\caption{The scale dependences of gravitational potential (top) and growth function (bottom) at the present epoch for dark energy with $w=const$ (left) and $c_a^2=0$ (right). Classical Lagrangian -- dotted line, tachyonic one -- solid.}\label{fig5}
\end{figure*}

The dependences of the gravitational potential and growth function on scale of perturbations for the present epoch are shown in Fig. \ref{fig5}. The presented plot supports the conclusion that the scale dependence in $w=const$-case is very weak and almost the same for both types of the Lagrangian (as mentioned above, a bit higher suppression of $g$ caused by classical field is due to the effective sound speed, which in this case is slightly larger than for tachyon). At small scales $\Psi$ and $g$ are practically scale-independent -- here the behavior is close to that in the ``quasi-$\Lambda$CDM'' model, i.e., model with $w=const\neq-1$ and without dark energy perturbations \cite{kul2008}. For $c_a^2=0$ the scale dependence for classical field is significantly stronger than for tachyonic one (in agreement with the result deduced from temporal dependences). The oscillations seen in the scale dependence of $\Psi$ for model with the tachyonic Lagrangian simply reflect the fact that small oscillations along the mean temporal dependence of gravitational potential come to $a=1$ for different $k$ in different phases, so they lie above or below the averaged curve.

In far future ($a\rightarrow\infty$) all perturbations, which in past have entered the particle horizon, will exit the event one. Both fields with $c_a^2=0$ tend to the cosmological term mimicry, therefore the perturbations will be entirely smoothed out. In $w=const$-case the situation is more complicated. When the inhomogeneities in dark energy are neglected, all perturbations will decay to 0. However, taking them into account leads to the freeze of gravitational potential after the event horizon exit, as it follows from the numerical analysis and asymptotic solutions. The dark energy perturbation variables $D^{(de)}$ and $V^{(de)}$ decay but $D_g^{(de)}$ and $D_s^{(de)}$ tend to the constant values. It should be mentioned that the latter variables decay in sub-horizon regime but after the event horizon leaving they start to grow in order to reach the frozen values. Nevertheless, the described features are irrelevant for choice of the best fitting dark energy model because they correspond to the future behavior of super-horizon and thus non-observable inhomogeneities.

\subsection{Transfer functions}

The observations of the large-scale structure of the Universe provide us with the power spectrum of matter density perturbations $P(k,a)$,
which is related to primordial (post-inflationary) one via the transfer function $T(k,a)$ as follows:
\begin{eqnarray*}
 P(k,a)=A_sk^{n_s}T^2(k,a),
\end{eqnarray*}
where $A_s$ is the normalization constant, $n_s$ -- scalar spectral index.

The transfer function can be built for each component of the Universe, and for dark energy too \cite{c_s^2_sp}. In this paper we are interested in scale dependences of the transfer functions
\begin{eqnarray}
 T_m(k,a)=\frac{D^{(m)}(k,a)}{D^{(m)}(k_{min},a)}\frac{D^{(m)}(k_{min},a_{min})}{D^{(m)}(k,a_{min})}                                \end{eqnarray}
and 
\begin{eqnarray}
 T_{de}(k,a)=\frac{D^{(de)}(k,a)}{D^{(de)}(k_{min},a)}\frac{D^{(de)}(k_{min},a_{min})}{D^{(de)}(k,a_{min})}.
\end{eqnarray}
Here $a_{min}$ is a very small scale factor, for which all scales of interest are still superhorizon. In Fig. \ref{fig6} these dependences are presented for the present epoch ($a=1$). It can be seen that in the $w=const$-case the scale dependence of $T_m^2$ is weak and both curves overlap. It is an illustration of the fact that the influence of perturbations of such fields on matter ones is almost negligible. For comparison we show the cold dark matter (CDM) transfer function computed for $w=const$ using CAMB\footnote{http://camb.info} \cite{camb}. Such calculation takes into account not only the gravitational coupling of the dark energy to pressureless matter but all physical processes governing the evolution of its perturbations. We see that the influence of dark energy inhomogeneities on CDM ones is almost negligible comparing to the effect caused by other components and their interactions.

In $c_a^2=0$-case the transfer functions decay more quickly than in $w=const$-one. This fact suggests that the influence of such fields on the computed by CAMB CDM transfer function (corresponding to the real Universe) could become comparable to the other processes determining it. It should be noted that in agreement with previous conclusions the perturbed classical field affects the matter transfer function more than tachyonic one.

In the bottom panel the scale dependence of $T_{de}^2$ is shown for the present epoch. As we can see, comparing with the dust matter transfer functions the dark energy ones are reduced by approximately 9 orders for both $w=const$-fields, 10 orders for the classical field with $c_a^2=0$ and 5 orders for $c_a^2=0$-tachyon at small scales ($0.1$ Mpc$^{-1}$).
The scale dependence of $T_{de}^2$ is monotonous for $w=const$-fields with both Lagrangians and $c_a^2=0$-classical field (the latter is smaller approximately by an order than the former ones). For tachyonic field the scale dependence of transfer function oscillates with decreasing amplitude. These oscillations are simply produced by different phases, which the density perturbations of different scales have at $a=1$.  It is interesting to see that their period (especially for large $k$) is almost constant and equals approximately $0.00045$ Mpc$^{-1}$.

\section{Conclusion}\label{concl}

The evolution of the scalar linear perturbations is studied for the 2-component (dust matter and minimally coupled dark energy) cosmological models. The dark energy component is assumed to be either classical or tachyonic scalar field with the potentials constructed for EoS parameters satisfying the condition either $w=const$ or $c_a^2=0$.

The analysis of gravitational instability of the single field models shows that the behavior of subhorizon perturbations is oscillatory in all studied cases. Thus the fields with $c_a^2=0$, although on the background level they mimic dust matter at early epoch and the cosmological constant in far future, can't serve as unified dark matter because they don't cluster on scales smaller than particle horizon.

\begin{figure}
\centerline{\includegraphics[width=5cm]{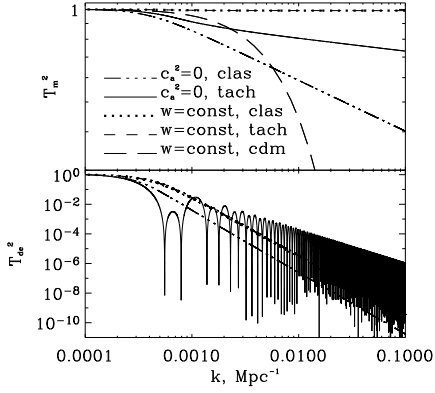}}
\caption{The transfer functions for dust matter (top) and dark energy (bottom) at $a=1$. The computed by CAMB cold dark matter transfer function for fields with $w=const$ is shown for comparison.}\label{fig6}
\end{figure}

In all 2-component models the dark energy is almost smoothed out on subhorizon scales. The perturbations of fields with $w=const$ grow or decay without visible oscillations and their evolution is similar for both types of Lagrangian (Fig. \ref{fig2}). The behavior of the $c_a^2=0$-tachyon density perturbations remains oscillatory (unlike that of the corresponding classical field, see Fig. \ref{fig3}). Such difference is caused by behavior of the effective sound speed of dark energy. In $w=const$-case the quantity $c_s^2$ is large and almost the same for both types of Lagrangian at all stages of evolution of perturbations as soon as the observational data prefer the values of the EoS parameter close to $-1$. In $c_a^2=0$-case the effective sound speed of classical field is always equal to $1$, while $c_s^2$ of tachyon varies in time as $-w$, so it is relatively small before the beginning of dark energy dominated epoch. It was found that the inhomogeneities of scalar fields with larger effective sound speed are more affected by $\rho_m\neq0$. This effect, though undetectable, could be interesting from the purely theoretical point of view as it might be useful for better understanding of the dark energy nature. The same applies to the dark energy transfer functions.

The caused by perturbed dark energy scale dependence of the nonrelativistic matter growth and transfer functions as well as of the gravitational potential (Fig. \ref{fig4}-\ref{fig6}) is small and almost the same for both fields with $w=const$. In $c_a^2=0$-case it is large and different for classical field and tachyon. Such behavior is due to the effective sound speed of dark energy. The discussed scale dependence provides us with simple tests for choice of model which describes the observational data in the best way. The tests can be based on measurements of the matter density perturbations power spectrum and late ISW-effect.

\begin{acknowledgments}
This work was supported by the project of Ministry of Education and Science of Ukraine (state registration number 0107U002062) and the research program of National Academy of Sciences of Ukraine (state registration number 0107U007279). O. S. thanks the SEMPER POLONIA Foundation for the financial assistance.
\end{acknowledgments}

\appendix

\section{Evolution of scalar field variables and their perturbations}\label{ap}
Evolution equations for the background scalar fields $\phi(\eta)$, $\xi(\eta)$ and their linear perturbations $\delta{\phi}(k,\eta)$, $\delta{\xi}(k,\eta)$ can be obtained either from the Lagrange-Euler equations or from the differential energy-momentum conservation law ${T^i_{0;i}}^{(de)}=0$.

In the discussed cases they are following:
\begin{eqnarray*}
&&\ddot{\phi}+2aH\dot{\phi}+a^2\frac{d{U}}{d\phi}=0,\\
&&\ddot{\delta{\phi}}+2aH\dot{\delta{\phi}}+\left[k^2+a^2\frac{d^2{U}}{d\phi^2}\right]\delta{\phi}+2a^2\frac{d{U}}{d\phi}\Psi
\nonumber\\
&&-4\dot{\Psi}\dot{\phi}=0
\end{eqnarray*}
for fields with Klein-Gordon (classical) Lagrangian and
\begin{eqnarray*}
&&\frac{\ddot{\xi}-aH\dot{\xi}}{1-\left(\dot{\xi}/a\right)^2}+3aH\dot{\xi}+a^2\frac{d\tilde{U}}{\tilde{U}d\xi}=0,\\
&&\ddot{\delta{\xi}}+\left[2aH-9aH\left(\frac{\dot{\xi}}{a}\right)^2-2\frac{d\tilde{U}}{\tilde{U}d\xi}\dot{\xi}\right]\dot{\delta{\xi}}+\left[k^2\right.\nonumber\\
&&\left.+a^2\left(\frac{d^2\tilde{U}}{\tilde{U}d\xi^2}-\left(\frac{d\tilde{U}}{\tilde{U}d\xi}\right)^2\right)\right]
\left(1-\left(\frac{\dot{\xi}}{a}\right)^2\right)\delta{\xi}\nonumber\\
&&-\dot{\Psi}\dot{\xi}-3\dot{\Psi}\dot{\xi}\left(1-\left(\frac{\dot{\xi}}{a}\right)^2\right)+2\Psi a^2\frac{d\tilde{U}}{\tilde{U}d\xi}\nonumber 
\\
&&+
6aH\Psi\dot{\xi}\left(\frac{\dot{\xi}}{a}\right)^2=0
\end{eqnarray*}
for fields with Dirac-Born-Infeld (tachyonic) one.

\end{document}